# A physics-based solver to improve the illumination of cylindrical targets using spherically-distributed high power laser systems.

P.-A. Gourdain, Extreme State Physics Laboratory, Department of Physics and Astronomy, University of Rochester, NY 14618, USA and Laboratory for Laser Energetics, University of Rochester, NY 14618, USA

Abstract

In recent years, our understanding of high energy density plasmas has played an important role in improving inertial fusion confinement and in emerging new fields of physics, such as laboratory astrophysics. Every new idea required developing innovative experimental platforms at high power laser facilities, such as OMEGA or NIF. These facilities, designed to focus all their beams onto spherical targets or hohlraum windows, are now required to shine them on more complex targets. While the pointing on planar geometries is relatively straightforward, it becomes problematic for cylindrical targets, or target with more complex geometries. This publication describes how the distribution of laser beams on a cylindrical target can be done simply by using a set of physical laws as a pointing procedure. The advantage of the method is threefold. First, it is straightforward, requiring no mathematical enterprise besides solving ordinary differential equations. Second, it will converge if a local optimum exists. Finally, it is computationally inexpensive. Experimental results show that this approach produces a geometrical beam distribution that yields cylindrically symmetric implosions.

## 1 Introduction

In recent years, a large fraction of high power laser shots was focused on research not directly related to inertial confinement fusion. For instance, laser facilities like OMEGA[1] or NIF[2] have been used successfully to produce and study plasma jets[3] or hydrodynamic shocks[4]. In these experiments, the target geometry was not spherical and beam positioning was non-trivial. In the case of a sphere, maximizing the laser intensity on target, distributing homogeneously the beams on target and minimizing incidence angles come down to pointing all beams at the center of the sphere. However, when a cylindrical target is used, the beam distribution becomes more problematic, especially when a 3-D thermal radiation code like VISRAD[5] has to be used to compute the absorption of light throughout a complex system of surfaces. One might think that the computational time would increase superpolynomially with the number of pointing locations. However, the locations on the target are not defined in advance since they are part of the problem we are trying to solve (unlike the traveling salesman problem where locations are given beforehand).

Global optimization problems are difficult to solve, especially when the space to explore is large. There are typically three different methods to search for an optimum. Deterministic methods use mathematical rigor to guarantee that a global optimum is found, given certain control parameters (e.g. linear programming[6], cutting plane methods[7]). Stochastic methods search the parameter using random processes (e.g. Monte-Carlo sampling[8]). Finally, heuristic approaches do partial searches of the parameter space which sufficient breadth that there is a large chance that the found optimum is a global one. In the last category, many algorithms physics-based solvers. One famous example is simulated annealing[9] where the amplitude of random perturbations between the different states of the problem is reduced after each iteration, ultimately freezing the system into a low energy state, like cooling would do to liquid metal. Another example is the evaporating lake algorithm[10] which found local optima before selecting from this set the one that is global. An error estimate is usually used to judge the quality of the solution. For the problem at hand, the error estimate requires the computation of the illumination of the surface using a thermal radiation code. This approach is computationally prohibitive and we decided to take a different approach. Rather than optimizing our beam distribution using an error estimate we decided to develop an algorithm that distributes the beam as evenly as necessary on the surface of the laser target and then ran a thermal radiation code to verify if the power is indeed evenly distributed. However, the convergence of the algorithm is guaranteed since the system of pointing location is just a mechanical system that will relax into an equilibrium located a minimum energy state. Experimental evidence presented in this paper shows that this simple approach yields well-balanced implosions and worked right out of the box.

While we focus on cylindrical target in this paper, this procedure can be extended to more complex 3D surfaces. Any surfaces can be used in principle, but practical considerations (e.g. shadowing) limit the applicability of the procedure. After introducing the theory behind the alignment procedure, this paper presents a series of illuminations for a cylinder using the OMEGA laser geometry and discusses the impact of the control parameters as well as initial conditions. In conclusion, we show that the procedure we developed produces cylindrical implosions.

## 2   The alignment procedure

We represent each pointing location on the target as a particle, which position is controlled by a simple physics solver. In the rest of this section will use the terms "electrons" or particles when talking about these pointing locations. Each laser beam of the device is assigned a given particle throughout the alignment process, which can travel freely on the target surface. The solver uses electrostatic repulsion to maximize the area illuminated by the beams and gravitational attraction to focus laser power near the mid-plane of the cylinder to maximize power density. Elastic forces (spring) control the incidence angle between the beam origin at the edge of the vacuum chamber and its pointing location on the target surface. Friction is added as a damping force to limit oscillations near the minimum (i.e. stable) state of energy. The location of the origin of the coordinate system should take advantage of target symmetries.

Electrostatic solvers (i.e. Riesz 2-energy) have been used successfully to distribute homogeneously points on sphere (i.e. spherical covering). This problem was also solved successfully using Riesz s-energy (s≠2)[11] or spiral approximation[12]. Most electrostatic solvers are expensive to run since they compute $N^2$ interactions, where N is the total number of points. A distance cut-off can be used to limit the interaction range which accelerates computations. However, this is not necessary when dealing with present day laser systems which have a limited number of beams. We use the following electrostatic force $\vec{E}_n$ on each pointing location

$$\vec{E}_n = \sum_{i \neq n} \frac{e^2}{4\pi\varepsilon_0} \frac{\vec{r}_n - \vec{r}_i}{|\vec{r}_n - \vec{r}_i|^3}, \tag{1}$$

where $e = 1.6 \times 10^{-19}$ C is the elementary charge, $\varepsilon_0$ is the vacuum permittivity, and $\vec{r}_n$ is the position vector of the $n^{th}$ electron.

Gravity is the only attractive force of this problem. The solver ignores the self-gravity of each electron. In our case the target has up-down symmetry and we can use the mid-plane as a gravitational attractor. An infinite plane with surface mass density $\sigma$ generates a gravitational force $\vec{G}_n$ on each electron given by

$$\vec{G}_n = -2\pi G \sigma m_e \frac{\vec{a}.\vec{r}_n}{|\vec{a}.\vec{r}_n|} \vec{a}, \tag{2}$$

where $G$ is the gravitational constant and $m_e = 9 \times 10^{-31}$ kg is the electron mass. $\vec{a}$ is the vector normal to the mid-plane. The ratio $\vec{a}.\vec{r}_n$ over $|\vec{a}.\vec{r}_n|$ is used to differentiate if the electron is "above" or "below" the mid-plane. So $G_n$ is set to zero when $|\vec{a}.\vec{r}_n| = 0$ rather than being undefined.

The elastic solver connects the entry point of each laser beam $\vec{A}_n$ at the edge of the vacuum chamber to its corresponding pointing location $\vec{r}_n$ on the target. We use a spring constant $k_n(\vec{r}_n, \vec{A}_n)$ for the elastic force

$$\vec{S}_n = k_n(\vec{r}_n, \vec{A}_n)(\vec{r}_n - \vec{A}_n). \tag{3}$$

The spring constant can be a function of the distance and incidence angle. We used herein the simplest setting $k_n(\vec{r}_n, \vec{A}_n) = k$ since our cylindrical surface is relatively easy to work with. Complex geometries may require spring constants which adapt to the local shape of the surface.

We now group all the forces to evolve the equation of motion along the target surface at the location $\vec{r}_n$

$$m_e \frac{d\vec{v}_n}{dt} = \vec{E}_n + \vec{G}_n + \vec{S}_n - \mu_n \vec{v}_n. \tag{4}$$

$\mu_n$ is the viscosity responsible for damping particle motion near equilibrium. We integrate this equation using a first order time integration scheme to get the particle velocity

$$\vec{v}_n(t + \Delta t) = (\vec{E}_n + \vec{G}_n + \vec{S}_n) \frac{\Delta t}{m_e} + (1 - \mu_n \Delta t) \vec{v}_n(t). \tag{5}$$

We find the new position of the particle by integrating the velocity

$$\vec{r}_n(t + \Delta t) = (\vec{v}_n(t + \Delta t) - [\vec{v}_n(t + \Delta t).\vec{N}_n]\vec{N}_n)\Delta t + \vec{r}_n(t). \tag{6}$$

where the unit normal to the surface at the location $\vec{r}_n$ is $\vec{N}_n$. This guarantees that the particle travels exclusively along the surface of the target. As we kept physical units in Eqs. (1) through (6), the time stepping is close to classical electron dynamics (~10-100 fs).

## 3 Illumination of a cylinder

We present in this section the configuration that was used for an experimental campaign at the OMEGA laser facility. 36 laser beams were used to implode a cylinder filled with argon gas. The cylinder was surrounded by two coils generating a magnetic field used to limit heat losses inside the compressed plasma. As a result, 24 beams of the OMEGA laser were blocked by the coil hardware and could not be used. The target was a paralene-N cylinder, 600 microns in diameter and 5 mm long. The thickness of the wall was 20 microns. We used the procedure described above to distribute the pointing locations onto the cylindrical target surface.

VISRAD computed the laser energy deposited using a Gaussian beam profile with 600 micron full-width half-max. We used the following parameters in the numerical procedure: $\Gamma=2\pi G\sigma=60\times10^9 m/s^2$, $k=10^{-18} N/m$, $\mu=5\times10^6 Ns/m$ and $\Delta t=2\times10^{-10} s$. The procedure yielded the distribution presented in Figure 1-a. The configuration "naturally" formed 4 rings of 9 beams around the target. The distance between the outer ring and the inner ring is larger than the distance between the two inner rings. This distribution gives a flat top to the intensity on the cylinder surface, presented in Figure 1-b. Figure 2-a shows the intensity profile at the azimuthal location $\theta=0$. The flat illumination region is 300 micrometer wide, which limits the effective implosion region to a sub-millimeter length. Figure 2-b shows the procedure kept the azimuthal intensity fluctuations below 1.7% at a given location (z=0 shown here). We then tested the beam distribution on the OMEGA laser. The 36 beams of the device were pointed to eight locations during the night shift before shot day. On shot day the 36 beams were moved to the final locations with a precision smaller than 10 microns using a gold sphere to verity experimentally pointing accuracy. Cylindrical convergence was obtained on the first shot. Figure 3 shows the implosion near stagnation.

To understand how the primary parameters (i.e. the electrostatic and gravitational forces) impact the procedure, we changed the strength of the gravitational force. Since Eq. (4) is defined up to a constant, the variation of one parameter is sufficient to highlight the mechanics of the procedure. Figure 4-a shows the pointing locations for three different $\Gamma$ factors. When this factor is too large, the pointing locations are grouped near the cylinder mid-plane. While the power density of this configuration is much larger than the optimum case of Figure 1, the area of the illuminated region is much smaller, as seen in Figure 4-b. Figure 4-c shows the case of Figure 1 for direct comparison, which has the widest illuminated region with acceptable local power fluctuations. When the $\Gamma$ factor is too small, Figure 4-a and d shows that a broadening of the illuminated area. While the power density is much lower, more pre-occupying is the loss of azimuthal symmetry of the illumination.

Since the electrostatic force is repulsive, the attractive force, which strength is proportional to the $\Gamma$ factor, controls the degree of packing of the pointing locations. At low factors, the pointing locations can occupy a large space. They are loosely packed. This explains the lack of order, as shown in Figure 4-a and d ($\Gamma=3\times10^{10}$). As the factor increases, the attractive force start to overcome the repulsive force and packing takes place and four groups (i.e. rings) of pointing locations have emerged. Figure 4-a and c ($\Gamma=6\times10^{10}$) show relatively ordered pointing locations in a crystal-like structure. This ordered configuration allows for homogeneous illumination with a wide illuminated area. When $\Gamma$ reaches $12\times10^{10}$, three rings form (see Figure 4-a and b). This configuration provides larger intensities. However, the illuminated area is now smaller. Increasing $\Gamma$ further would yield two-ring then one-ring configurations. Here again the intensity would increase while the illuminated area would decrease. Depending on the experimental goal, these configurations may be better (if large compressions are required) or worse (if large compressed volumes are required). When using 36 pointing locations, the 4-ring, 3-ring, 2-ring and 1-ring systems always yield quasi-axi-symmetric configurations. Using more (or less) pointing locations would generate rings with a different number of points, in turn yielding poor illumination.

Since the spring force is not dominant, all the pointing locations are equivalent to first order. The locations closest to the mid plane at the beginning of the procedure will still be the closest to the mid-plane when the procedure has converged. As a result, the beam order at the beginning will be the same at the end, providing we assign one beam to each pointing locations and this assignment does not change during the whole procedure. Therefore, the final locations strongly depend on the initial conditions (which beam is assigned to which pointing location). For instance, if our goal is to have the largest possible illuminated area that a particular configuration can allow, then having large incident angles will help keeping the illuminated area as large as possible (albeit reducing intensity). In previous cases, we had assigned beams that were the furthest away from the cylinder mid-plane to pointing locations that were closest to the mid-plane. Therefore, the optimized configuration had beams at large incidence angles. Now if we switch the assignment then the optimized configuration has beams at a smaller angle of incidence, de facto reducing the illuminated area, as shown in Figure 5. . This is equivalent to switching the inner and outer rings shown in Figure 1-a.

This procedure described in this paper converges is less than 40,000 thousand steps. It takes less than 10s to run on a modern laptop, using no parallel computations. This procedure can be accelerated if one is interested at scanning a given parameter space. After the initial procedure has converged, it can be used as input for the next search with one of the parameters slightly altered, requiring an order of magnitude less steps to converge.

## 4 Discussion

The robustness of the procedure herein can be extended to further tune the beam distribution. It is possible to alter Eqs. (1) and (2) to follow experimental imperatives. For instance, one can redefine Eq. (1) to take into account the beam cross-section on target. If a laser beam with circular cross-section hits a target at an oblique angle then the beam cross-section on that target is elliptical rather than circular. Providing that the average strength of the field $\vec{G}_n$ is similar to the one computed in Eq. (1) then the procedure will be stable. Further if the overall illumination profile requires controlling the drop off of the illumination profile shown in Figure 2-a different electrostatic and gravitational laws could be used:

$$\vec{G}_n = -2\pi G \sigma m_e \frac{\vec{r}_n - \vec{r}_i}{|\vec{r}_n - \vec{r}_i|^m}. \qquad (7)$$

where $m$ is an integer. Replacing Eq. (2) with Eq. (7) in the procedure will broaden the illumination profile as $m$ increases. As discussed earlier, if the average strength of $\vec{G}_n$ is similar for Eq. (2) and (7), then the procedure will converge. As an example we implemented an strategy with minimize the spread of incidence angles using the following elastic solver

$$\vec{K}_n = L_n \vec{B}_n \cdot \vec{T}_n (\theta_n - \theta_{ave}) \vec{T}_n. \qquad (8)$$

Here $\theta_n$ is the incident angle of the laser with direction $\vec{B}_n$, $\theta_{ave}$ is the average incident angle. $\vec{T}_n$ is the unity vector tangent to the target at the $n^{th}$ location and perpendicular to the cylinder axis. Without incidence angle correction, the minimum incidence angle is 33.87° and the maximum angle is 42.64° for the beam distribution generating the illumination given in Figure 5. The angle spread is close to 10°. When the procedure presented above is used, the minimum incident angle is 40.09° and the maximum angle is 42.86°. The angle spread is now reduced to 3°. The difference in the two distributions is shown in Figure 6-a. The area of each circle corresponds to a beam number and is completely arbitrary. It is only used to show how the beams were relocated in the second configuration. The illuminations shown in Figure 5 and Figure 6-b are similar which shows that illuminations which optimize power density together with illuminated area can also accommodate incidence angle constraints.

# 5  Conclusions

This paper has shown that multi-laser high power beam systems can be turned into versatile high energy density plasma drivers. We used a heuristic procedure based on physical laws to distribute laser beams on a target surface as compactly as required by experimental imperatives. To implement this procedure, we have replaced the pointing position on the target by virtual particles with physical properties (like mass and charge). For simple targets (cylinder or foil), physical electrostatic, gravitational and elastic forces distribute the pointing location of each laser beam in such a way that the illumination is quasi-homogenous. These rules can be modified to accommodate for more complex shapes to give the geometrical locations of pointing locations. As an example, we demonstrated experimentally that one could obtain excellent axi-symmetric convergence for cylindrical targets on OMEGA in one single shot, using this procedure. By coupling this procedure to a 3-D thermal radiation code such as VISRAD, very little work was required on shot day to tune the beam power balance. The implementation of an automated interface between the tuning code and VISRAD has allowed fast turnaround time to investigate hundreds of configurations. The procedure presented herein will enable high power lasers to tackle more complex geometries, turning them into more flexible experimental platforms to study matter under extreme conditions.

**Acknowledgements**: This research was supported by the DOE grant number DE-SC0016252.

[1] T. R. Boehly, T. R. Boehly, D. L. Brown, R. S. Craxton, R. L. Keck, J. P. Knauer, J. H. Kelly, T. J. Kessler, S. A. Kumpan, S. J. Loucks, S. A. Letzring, F. J. Marshall, R. L. McCrory, S. F. B. Morse, W. Seka, J. M. Soures, and C. P. Verdon, Opt. Commun. **133**, 495 (1997)

[2] J. D. Lindl and E. I. Moses, Phys. Plasmas **18**, 050901 (2011)

[3] P. Hartigan, J. Foster, B. Wilde, R. Coker, P. Rosen, J. Hansen, B. Blue, R. Williams, R. Carver and A. Frank, ApJ **705**, 1073 (2009)

[4] A.B. Reighard, R.P. Drake, J.E. Mucino, J.P. Knauer, M. Busquet, "Planar radiative shock experiments and their comparison to simulations", *Physics of Plasmas* **14**, 056504 (2007)

[5] http://www.prism-cs.com/Software/VisRad/VisRad.htm

[6] D. G. Luenberger, Y. Ye, Linear and non-linear programming, Springer 4th Ed. (2008)

[7] H. Tuy, Cutting Plane Methods for Global Optimization, Encyclopedia of Optimization, Series Ed. C. A. Floudas, P. M. Pardalos, p. 590, Springer (2009)

[8] B. Hesselbo and R. B. Stinchcombe, Phys. Rev. Lett. 74, 2151 (1995)

[9] S. Kirkpatrick; C. D. Gelatt; M. P. Vecchi, Science, New Series, Vol. 220, No. 4598, 671 (1983).

[10] P.-A. Gourdain, J.-N. Leboeuf, R.Y. Neches, J. Comp. Phys. **216**, 275 (2006)

[11] D. P. Hardin, E. B. Saff, Notices of the AMS **51**, 1186 (2004)

[12] E. B. Saff and A. B. J. Kuijlaars, Mathematical Intelligencer **19**,5 (1997)

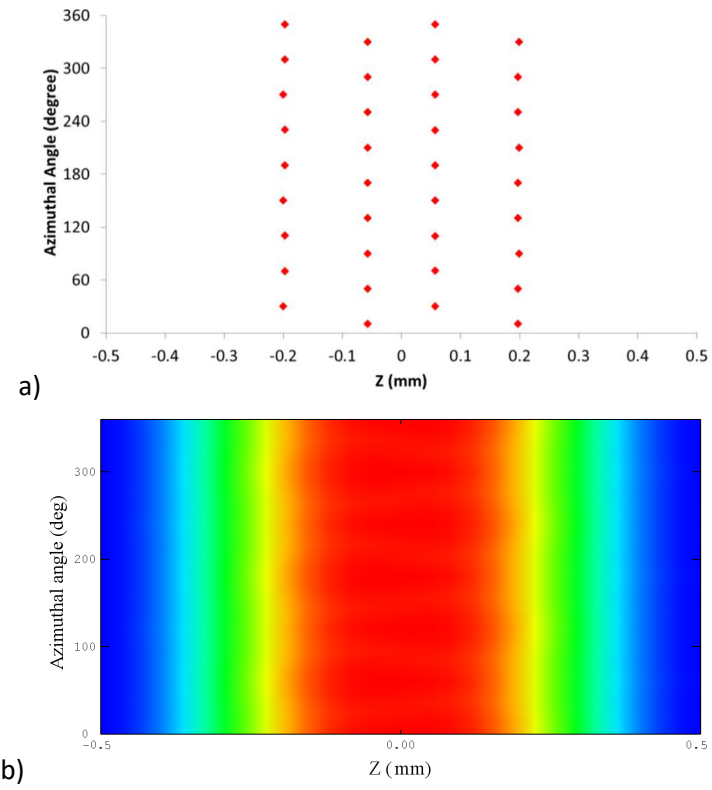

**Figure 1. a) Pointing locations on the cylinder surface and b) intensity on the target surface using arbitrary units computed by VISRAD**

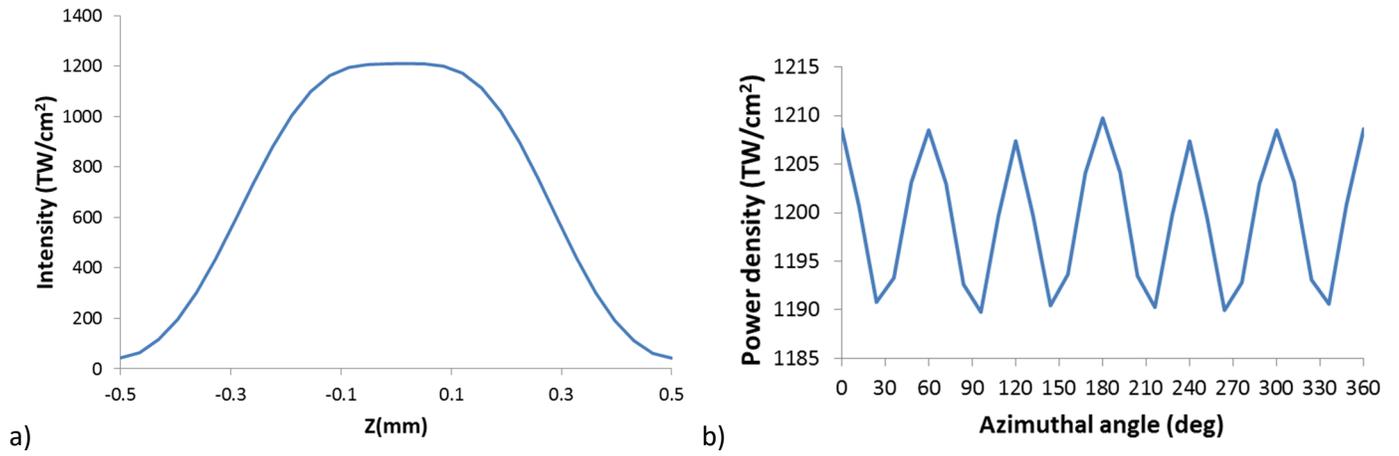

Figure 2. a) Longitudinal absolute intensities (profile at θ=0) and b) azimuthal absolute intensities (at z=0) computed by VISRAD

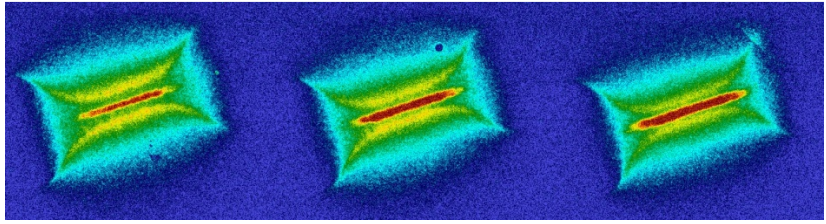

Figure 3. X-ray framing camera showing the self-illumination of the implosion near stagnation. The first picture was taken 1.7ns after the lasers hit the target. The next pictures were taken at 100 ps time intervals.

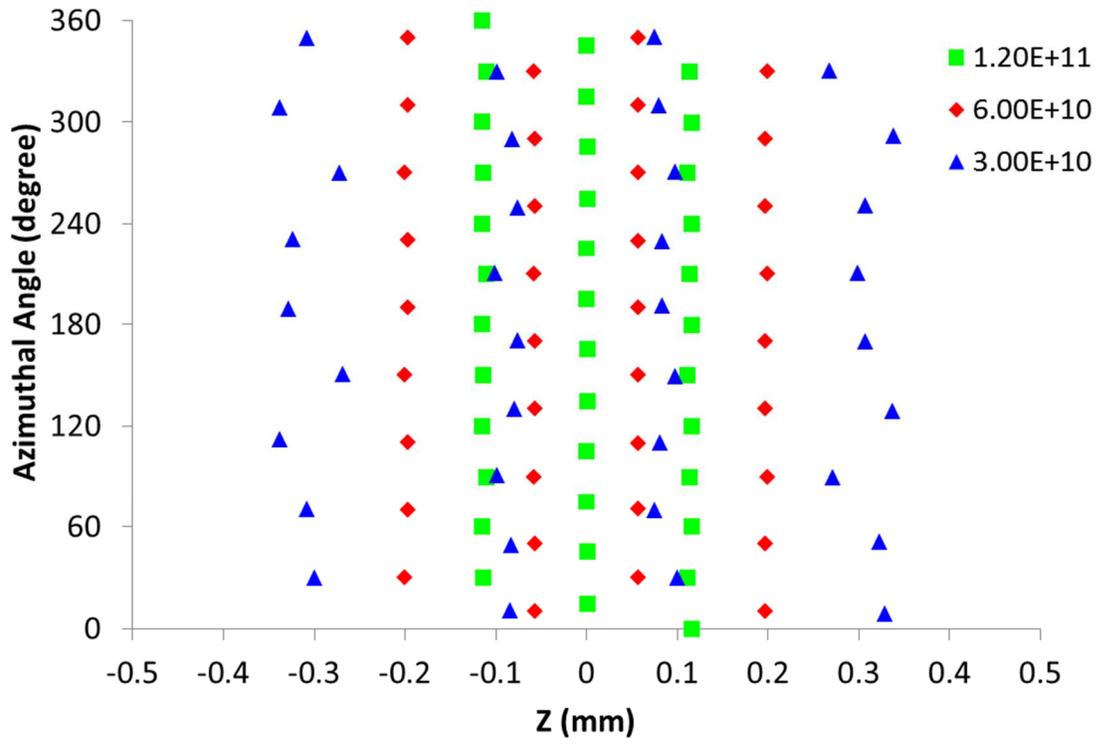

a)

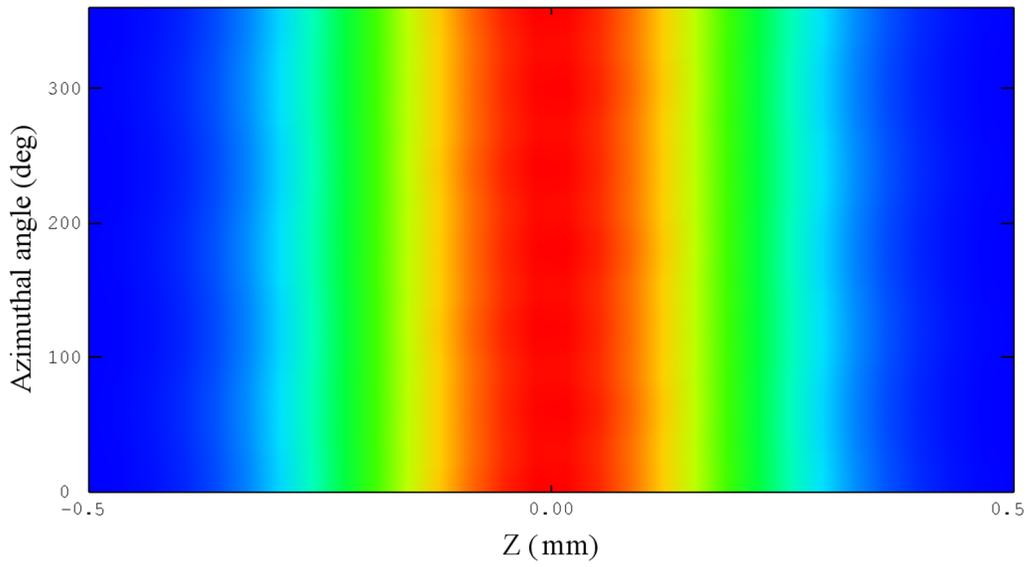

b)

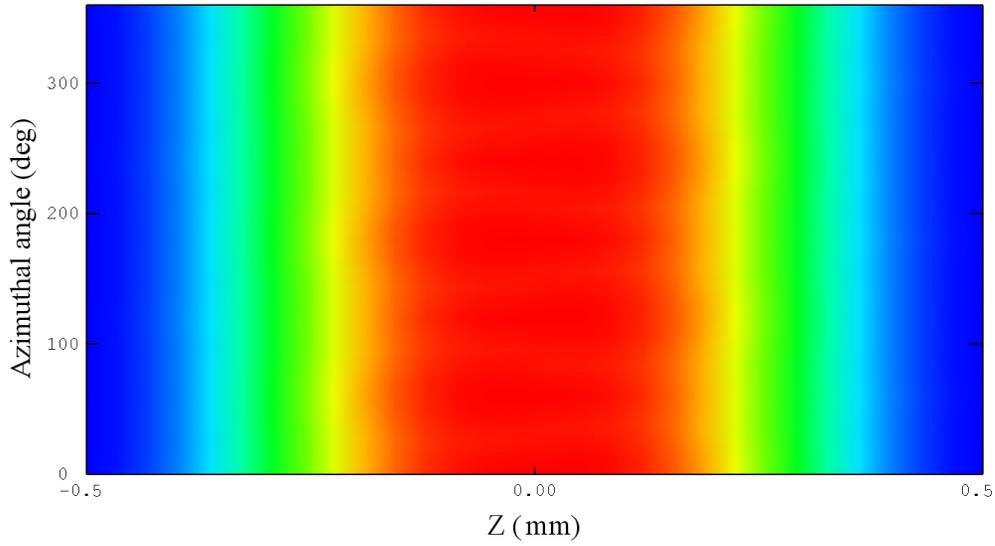

c)

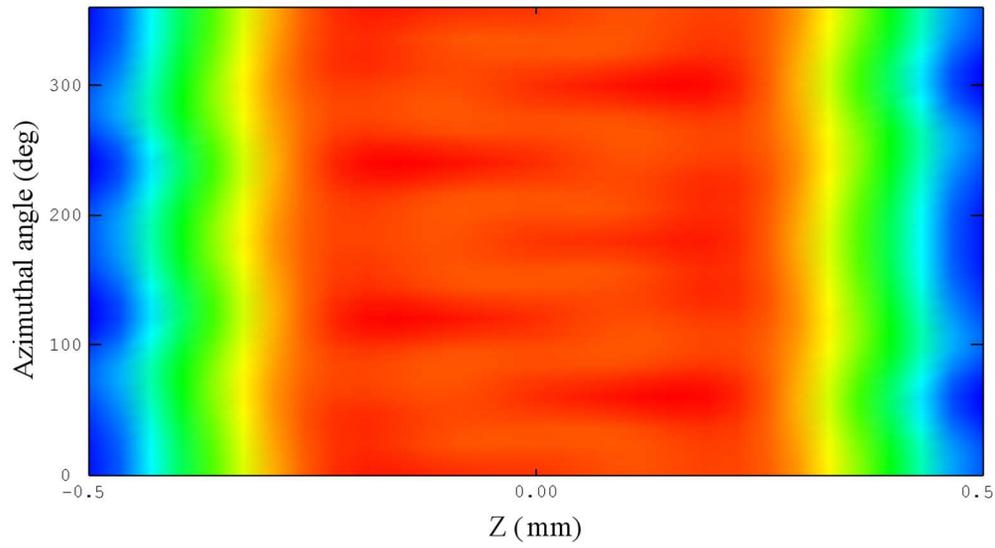

d)

**Figure 4. a) Pointing distribution for a factor Γ=1.2E11 (squares), 6E10 (diamonds) and 3e10 (triangles) and the resulting Visrad computation for b) Γ=1.2E11, c) Γ=6E10 and d) Γ=3E10 using arbitrary units. The color scale for all three panels has be rescaled from 0 to 1.**

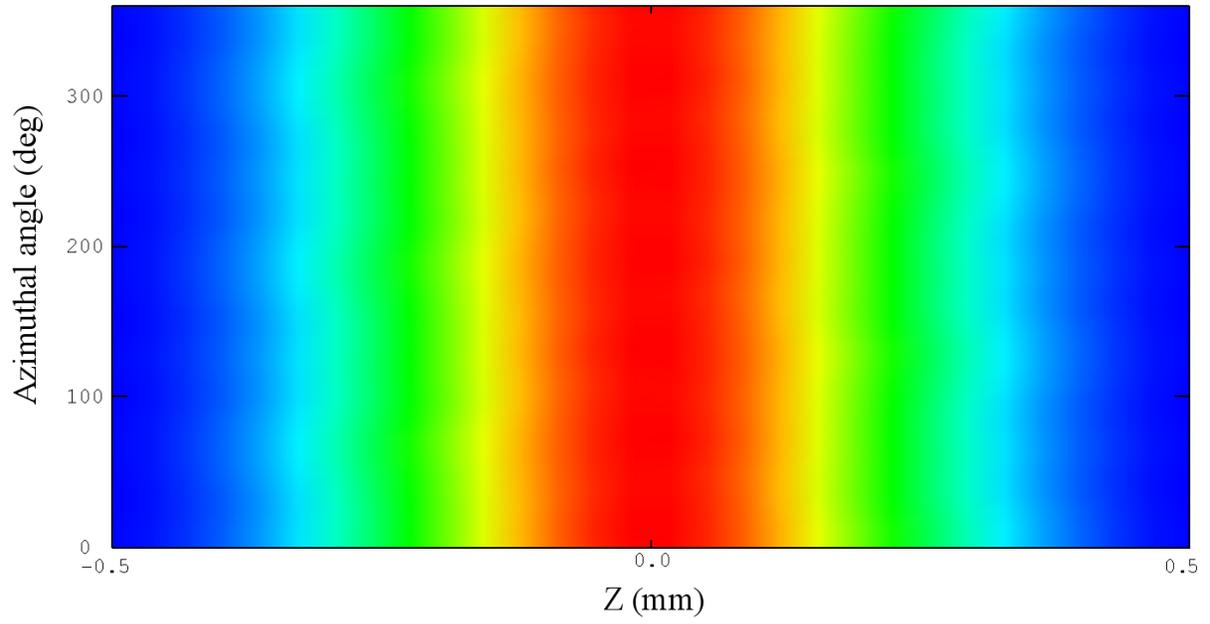

**Figure 5. Intensity (A.U.) computed by VISRAD when the beams have a smaller incident angle compared to the configuration of Figure 1. The actual distribution of the pointing location on the cylinder surface is identical to the one shown in Figure 1-a.**

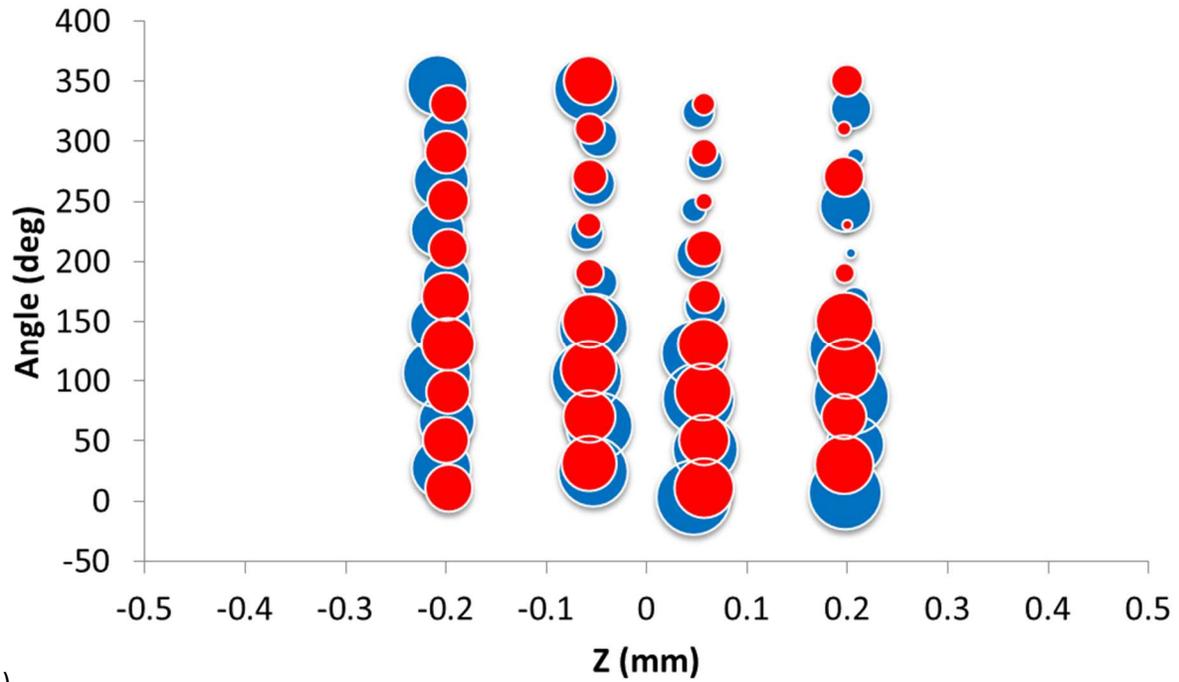

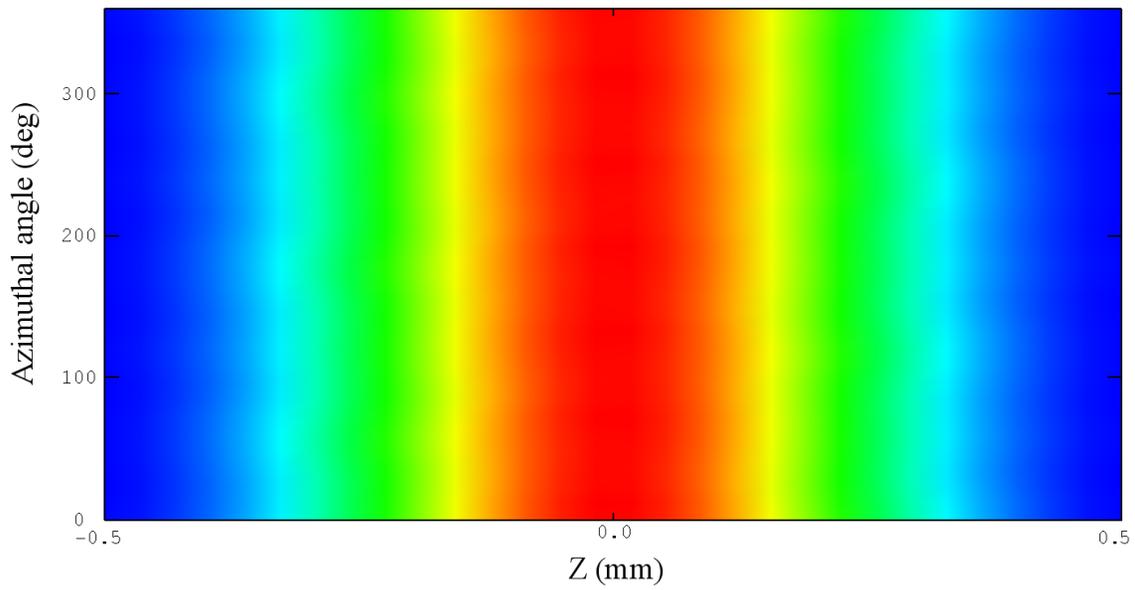

Figure 6. a) Beam distribution for the optimized case of Figure 5 (red circles) with no incidence angle optimization and with incidence angle optimization (blue circles). The area of each circle corresponds to an OMEGA beam number not to the size of the focal spot. b) Intensity (A.U.) computed by VISRAD for the optimization with incidence angle constraint.